\documentclass{aa}
\usepackage{graphics,epsfig}
\begin{document}

\title{Modelling the Milky Way through adiabatic compression of cold dark matter halo}

\author{V. F. Cardone\inst{1}
        \and M. Sereno\inst{2,3}}

\offprints{winny@na.infn.it}

\institute{Dipartimento di Fisica ``E.R. Caianiello'', Universit\`{a} di Salerno, and INFN, Sez. di Napoli, Gruppo Coll. di Salerno, Via S. Allende, 84081 - Baronissi (Salerno), Italy
\and Dipartimento di Scienze Fisiche, Universit\`{a} di Napoli, and INFN, Sez. di Napoli, Complesso Universitario di Monte S. Angelo, Via Cinthia, 80126 Napoli, Italy
\and Istituto Nazionale di Astrofisica - Osservatorio Astronomico di Capodimonte, Salita Moiariello 16, 80131 Napoli, Italy}

\date{Received / Accepted }

\abstract{We use the adiabatic compression theory to build a physically
well\,-\,motivated Milky Way mass model in agreement with the observational
data. The visible mass of the Galaxy is distributed in a spheroidal bulge
and a multi\,-\,components disc parametrized by three galactic parameters,
the Sun distance to the galactic centre, $R_0$, the total bulge mass,
$M_\mathrm{bulge}$, and the local disc surface density, $\Sigma_{\odot}$.
To model the dark matter component, we adiabatically compress a Navarro,
Frenk and White (NFW) halo (with concentration $c$ and total mass
$M_\mathrm{vir}$) for fixed values of the spin parameter, $\lambda$, the
fraction of the mass in baryons, $m_\mathrm{b}$, and the thin disc
contribution to total angular momentum, $j_\mathrm{d}$. An iterative
selection procedure is used to explore in very detail the wide space of
parameters only selecting those combinations of $\left\{ R_0,
M_\mathrm{bulge},
\Sigma_{\odot},
\lambda, m_\mathrm{b}, j_\mathrm{d}, c, M_\mathrm{vir} \right\}$ that give rise to a Milky Way
model in agreement with the observational constraints. This analysis
leads us to conclude that only models with $R_0 = 8.5$\,kpc, $0.8 {\times}
10^{10}
\ M_{\odot} < M_\mathrm{bulge} < 1.6 {\times} 10^{10} \ M_{\odot}$ and $49 \ M_{\odot} \ pc^{-2} \le
\Sigma_{\odot} \le 56 \ M_{\odot} \ pc^{-2}$ can be reconciled with the set of
observational constraints. As regard the parameters entering the
adiabatic compression, we find $0.03 \le \lambda \le 0.10$ and $0.04
\le m_\mathrm{b} \le 0.10$, while the final estimates of the parameters
describing the initial halo profile turn out to be $5
\stackrel{<}{\sim} c \stackrel{<}{\sim} 12$ and $7 {\times} 10^{11} \ M_{\odot}
\stackrel{<}{\sim}
M_\mathrm{vir} \stackrel{<}{\sim} 17 {\times} 10^{11}
\ M_{\odot}$ (all at 95.7$\%$ CL).

\keywords{Galaxy: kinematics and dynamics -- Galaxy: structure -- galaxies: formation -- dark matter}}

\titlerunning{Modeling the Milky Way}

\maketitle

\section{Introduction}

The determination of the mass distribution of the Milky Way is a
classical task of astronomy (\cite{sch56,ca+os81,DB98}). The usual
framework for the origin of structures in the Universe is provided by
the Cold Dark Matter (CDM) paradigm. This standard cosmological theory
has proved very successful on large scale, in explaining both the
abundance and clustering of galaxies (\cite{pea+al01,ver+al01}) and
the power spectrum of the cosmic microwave background anisotropies
(\cite{deb+al00}), but is experiencing a number of difficulties on the
scales of galaxies and dwarf galaxies. CDM paradigm predicts an
over-abundance of satellites around the Milky Way and M31 by an order
of magnitude (\cite{Ketal99,moo+al99}). While the presence of a
photoionizing background can solve this ``sub-structure" problem
(\cite{som01}), difficulty about density profile still remains. Dark
matter dominated objects, such as dwarf and low surface brightness
galaxies, show rotation curves inconsistent with the central density
cusp predicted by CDM cosmology (\cite{mc+de98,ala+al01}).

The understanding of galaxy formation also remains unsolved in the
standard hierarchical model. Dissipationless CDM haloes are assumed to
form bottom-up via gravitational amplification of initial density
fluctuations. Gas carried with such haloes cools and contracts within
them to form luminous, independent self-gravitating units, which can
form stars, at the halo centres (\cite{fa+ef80,Betal86,MMW98}). The
halo profile affects both gas cooling and infall since it determines
the structural properties of the resultant heavy discs and dense
nuclear bulges. Even if the growth of dark haloes is not much affected
by the baryonic components, the halo gravitationally responds to the
dissipative baryonic infall and the present day CDM haloes can have
density profiles quite different from the original CDM prediction.

In this context, we want to address the density profile problem by
fitting CDM models to the observed properties of the Milky Way. The
Milky Way seems to be a typical system on a mass scale of
$10^{12}$\,M$_\odot$, mostly contributed by exotic particles, such as
weakly interacting massive particles or axions. Direct searches for
dark compact objects, such as MACHOs, in the Milky Way halo have been
performed by the MACHO and EROS collaborations through microlensing
surveys. According to the MACHO group (\cite{alc+al00}), the most
likely halo fraction in form of compact objects with a mass in the
range $0.1-1$ M$_\odot$ is of about $20\%$; the EROS collaboration
(\cite{las+al00}) has set a $95\%$ confidence limit that objects less
than $1$ M$_\odot$ contribute less than $40\%$ of the dark halo. These
upper limits on the fraction of compact objects are evidences
favouring that Galaxy halo can be described by the standard
cosmological approach.

The very detailed data, obtained from many independent techniques,
which characterize the Milky Way, make it a unique testing for
theories of galactic structure and baryonic infall. Actually, this
consideration has motivated Klypin et al. (2002) to perform an
analysis similar to the one we present in this paper. Although
conceptually analogous to that of Klypin et al. (2002), the procedure
we will use is much more detailed and allows to deeply explore the
whole parameter space. As a result, we will be able not only to
investigate the viability of the standard CDM paradigm when applied to
the Galaxy, but also to constrain a set of parameters that are, on the
contrary, held fixed in the work of Klypin et al. (2002).

The paper is organized as follows. In Sect. 2 we describe the
ingredients needed for our analysis and in Sect.~\ref{adia} we
introduce briefly the adiabatic compression theory. Section 4 is
devoted to the description of the observational constraints we have
for the Milky Way, while in Sect. 5 we explain how we investigate the
parameter space to select only models in agreement with the data. The
impact of the different selection criteria on the parameter space is
discussed in Sect. 6 and the results of our analysis are reported in
Sect. 7. Section 8 is devoted to some final considerations.

\section{Model ingredients}

Traditional models of spiral galaxies usually include at least three
dynamical components\,: a spheroidal bulge, an exponential disc and a
dark halo. The lack of observational data precludes from using more
detailed multi\,-\,component mass models since it is not possible to
break the degeneracy among the many parameters involved. However, due
to our privileged position, cinematic and photometric data on the
Milky Way are numerous enough to require the use of a
multi\,-\,component model in order to have a unified picture. To this
aim, we model the mass distribution in the Milky Way introducing five
components (the bulge, the thin and thick discs, the interstellar
medium disc and the dark halo) that we describe in detail in the
following subsections.

\subsection{The bulge}

The morphology of the Galactic bulge (defined as the spheroid within
the galactic coordinates $|l| < 20^o$ and $|b| < 10^o$ region around
the Galactic centre) is much harder to ascertain than that of the
bulges in many external galaxies, because of obscuration by
interstellar dust due to our position in the Galactic plane. In the
recent years, however, striking images of the Galactic bulge have been
obtained in the near infrared (at wavelengths of $1.25$, $2.2$, $3.5$
and $4.9 \
\mu m$) by the DIRBE experiment on board the COBE satellite
(\cite{Aetal94,Wetal94}) allowing to study with good accuracy its
structure. These images suggest that the stellar distribution in the
bulge is bar\,-\,shaped, i.e. that the bulge is not rotationally
symmetric. However, a bar\,-\,like structure can not be used since the
standard adiabatic compression theory formally assumes that all the
mass distributions have spherical symmetry. On the other hand, it is
possible to achieve a relatively good agreement with the near infrared
photometric data also using spheroidal models for the bulge
(\cite{KDF91,D95}). Thus, we will describe the bulge as a spheroidal
density distribution (\cite{DB98})\,:

\begin{equation}
\rho_\mathrm{b} = \rho_{0} \left ( \frac{m}{r_{0}} \right )^{\gamma}
\left ( 1 +  \frac{m}{r_{0}} \right )^{\gamma - \beta} \ {\rm e}^{-m^2/r_{t}^{2}} \ ,
\label{eq: rhobulge}
\end{equation}
where
\begin{equation}
m^2 \equiv ( R^2 + z^2/q^2 )^{1/2} \ ,
\label{eq: mdef}
\end{equation}
with $R$ the galactocentric radius and $z$ the height above the
equatorial plane. Thus the density of the bulge is proportional to
$r^{-\gamma}$ for $r << r_{0}$, to $r^{-\beta}$ for $r_{0} << r <<
r_\mathrm{t}$ and softly truncated at $ r = r_\mathrm{t}$. Fitting of
the model to the observed infrared photometric {\it COBE/DIRBE} data
yields the values of four of the five bulge parameters
(\cite{DB98})\,:
\begin{displaymath}
\beta = \gamma = 1.8 \ , \ q = 0.6 \ ,
\ r_{0} = 1 \ {\rm kpc} \ , \ r_\mathrm{t} = 1.9 \ {\rm kpc} \ .
\end{displaymath}

The density normalization $\rho_{0}$ is not determined from the
fitting relation, but it is easily related to the total mass of the
bulge, $M_\mathrm{bulge}$. Integrating Eq.(\ref{eq: rhobulge}), one
gets
\begin{equation}
M_\mathrm{b}(m) = 0.518 M_\mathrm{bulge} m^{1.2} \ _1F_1[0.6, 1.6,
-0.27 m^2]
\label{eq: truemassbulge}
\end{equation}
where $_1F_1$ is the hypergeometric function and $M_\mathrm{bulge}$ is
related to the central density $\rho_{0}$ (in $M_{\odot}/pc^3$) as\,:

\begin{equation}
M_\mathrm{bulge} = 1.60851 {\times} 4 \pi q \rho_{0} \ .
\label{eq: rhomtot}
\end{equation}

In the adiabatic compression formalism, all the galaxy components are
assumed to be spherical. To this aim, it is needed to describe the
bulge mass distribution with a modified ``spherical" version of
Eq.(\ref{eq: truemassbulge}). A simple and reasonable way to solve
this problem is to substitute Eq.(\ref{eq: truemassbulge}) with the
following one\,:

\begin{equation}
M_\mathrm{b}^\mathrm{sph}(r) \simeq M_\mathrm{bulge} f(r) \ ,
\label{eq: sphermassbulge}
\end{equation}

\begin{displaymath}
f(r) = 1 - {\rm e}^{-1.867 r} (1 + 1.543 r + 0.1898 r^2 +
\end{displaymath}
\begin{equation}
\ \ \ \ \ \ \ \ \ \
0.6349 r^3 - 0.6109 r^4 +0.1491 r^5 - 0.01126 r^6 ) \ ,
\label{eq: frapprox}
\end{equation}
where now $r$ is the usual spherical radius. The spherical mass
distribution in Eq.({\ref{eq: sphermassbulge}) has been defined so
that the mean density inside the spherical radius $r$ is the same as
the one within the elliptical radius $m$, $M_\mathrm{b}(m)$, i.e.\,:

\begin{displaymath}
\frac{M_\mathrm{b}^\mathrm{sph}(r)}{4/3 \ \pi \ r^3} = \frac{M_\mathrm{b}(m)}{4/3 \ \pi \ q \ m^3} \ .
\end{displaymath}
Using Eq.(\ref{eq: sphermassbulge}) instead of Eq.(\ref{eq:
truemassbulge}) introduces a systematic error when describing the very
inner regions of the Galaxy. Actually, this error is expected to be
not a serious one since the bulge contribution to the total mass
budget of the Galaxy is indeed small. Moreover, as we will see later,
most of the observational constraints we will use probe a region of
the Galaxy that is so far from the inner bulge dominated region that
its dynamical effect could be described even by modelling this
component as a pointlike mass. However, to further reduce this effect,
we will use Eq.(\ref{eq: sphermassbulge}) for the bulge mass
distribution, but we consider the exact rotation curve
(\cite{BT87})\,:

\begin{equation}
v_\mathrm{c}^{2}(R) = 4 \pi G q \int_{0}^{R}
{\frac{\rho_\mathrm{b}(m^2) m^2 dm}{\sqrt{R^2 - (1 - q^2) m^2}}} \ .
\label{eq: vrotbulge}
\end{equation}
To fully characterize the bulge we only need its total mass,
$M_\mathrm{bulge}$. Dwek et al. (1995) found $M_\mathrm{bulge} = (1.3
{\pm} 0.5) {\times} 10^{10} \ M_{\odot}$.

\subsection{The disc}

Contrary to the bulge, the structure of the Milky Way disc is quite
easy to investigate given the large amount of available data. Disc is
made up of three components, namely the thin and the thick stellar
discs and the interstellar medium (ISM) disc. We model the two stellar
discs with the usual double exponential profile
(\cite{Freeman,DB98})\,:

\begin{equation}
\rho_\mathrm{d}(R,z) = \frac{\Sigma_\mathrm{d}}{2 z_\mathrm{d}} \
\exp{\left ( - \frac{R}{R_\mathrm{d}} - \frac{|z|}{z_\mathrm{d}} \right )}  \ .
\label{eq: discrho}
\end{equation}
The total mass of each stellar disc is $M_\mathrm{d} = 2 \pi
\Sigma_\mathrm{d} R_\mathrm{d}^{2}$. For the adiabatic compression, we
will adopt a 3-D mass distribution such that\,:

\begin{equation}
\label{eq: diskmass}
M_\mathrm{d}(r) = 2 \pi \int_{-\infty}^{+\infty}
\int_0^r\rho_\mathrm{d}(R,z)R dR dz \ .
\end{equation}
In Eq.(\ref{eq: diskmass}), $M_\mathrm{d}(r)$ has spherical symmetry
which is not formally correct since the disc is a highly flattened
structure. Actually, this approximation introduces a negligible error
as it is witnessed by the good agreement found between the predictions
of the adiabatic compression theory and the numerical simulations
(\cite{Jesetal02}). However, to evaluate the disc rotation curve and
the vertical force we use the original flattened density distribution
given in Eq.(\ref{eq: discrho}).\footnote{To this aim we use a C++
code kindly provided us by W. Dehnen which implements a modified
multipole technique developed by Kuijken \& Dubinski (1994, see also
Dehnen \& Binney 1998).}

Although the simple formula in Eq.(\ref{eq: discrho}) may fit well the
large-scale structure of the stellar discs, it is not able to
reproduce the smaller scale density fluctuations, which are prevalent
in the ISM component. Dame et al. (1987) have shown that there is very
little interstellar matter between the nuclear disc at 200\,pc and the
molecular ring at $R = 4 - 5$\,kpc. These local fluctuations strongly
affect the estimation of the Oort constants. Hence, we do not assume
an analytical expression for the ISM disc density profile, but we use
a third order polynomial interpolation of the data in Table D1 of
Olling \& Merrifield (2001) also including a 23.8$\%$ helium
contribution by mass. The 3-D mass distribution of the ISM disc is
then evaluated as in Eq.(\ref{eq: diskmass}). The ISM disc rotation
curve has been evaluated following the method described in Kochanek
(2002).

To fully characterize the disc model, we have to fix the geometrical
parameters $(R_\mathrm{d}, z_\mathrm{d})$ and a value for the central
(or the local) surface density. We fix the geometry of our stellar
discs giving their scale\,-\,length and scale\,-\,height as follows
(\cite{DB98})\,:

\begin{displaymath}
{\rm thin \ disc} : R_\mathrm{d} = \kappa R_0 \ {\rm kpc} \ , \
z_\mathrm{d} = 180 \ {\rm pc} \ ;
\end{displaymath}
\begin{displaymath}
{\rm thick \ disc} : R_\mathrm{d} = \kappa  R_0 \ {\rm kpc} \ , \
z_\mathrm{d}
= 1000 \ {\rm pc} \ .
\end{displaymath}
where $R_0$ is the distance of the Sun to the galactic centre and
$\kappa = (0.30 {\pm} 0.05)$ a scaling constant (\cite{DB98}). The exact
value of $R_0$ is quite uncertain, with most of the estimates ranging
from 7.0 to 8.5\,kpc (\cite{KL86,R93,OM00,OM01}). Given the importance
of this parameter in the modeling of the Galaxy, we will explore
models with different values of $R_0$.

We use the following estimate for the local surface density
(\cite{KG89})\,:

\begin{displaymath}
\Sigma_{\odot} = (48 \ {\pm} 8) \ M_{\odot} \ {\rm pc}^{-2} \ .
\end{displaymath}
$\Sigma_{\odot}$ accounts for the contribution of the discs (both
stellar and ISM), not taking into account the halo contribution.  The
total mass of each sub\,-\,disc is fixed specifying the fractional
contribution of each one to $\Sigma_{\odot}$. Using the relations
given in Dehnen \& Binney (1998), we fix:

\begin{displaymath}
\Sigma_\mathrm{thin}(R_0) = 14 \ \Sigma_\mathrm{thick}(R_0) \ , \
\end{displaymath}
while, for the ISM disc, the adopted value is (\cite{OM01})\,:

\begin{displaymath}
\Sigma_\mathrm{ISM}(R_0) = 14.5 \ M_{\odot} \ {\rm pc^{-2}} \ .
\end{displaymath}
It is noteworthy that there are also other estimates of
$\Sigma_{\odot}$ significantly lower than the one used here
(\cite{OM01,Ger02}).

\subsection{The halo}

Observational data may be fitted by a wide range of models, even
unphysical ones. This is why there are a lot of different dark halo
models which are claimed to describe well the density profile of this
component. To obtain physically interesting models, it is thus
important to impose constraints based on a physical theory of halo
formation and to select models which are both compatible with the data
and also physically well motivated. From this point of view, numerical
simulations of galaxy formation in hierarchical CDM scenarios are very
helpful since they predict the initial shape of the dark matter
distribution. In this paper, we assume a NFW profile (\cite{NFW97}) as
initial dark matter halo. The main properties of the NFW model are\,:

\begin{equation}
\rho(r) \equiv \frac{\rho_s}{x \ (1+x)^2} \ , \ x = r/r_s
\label{eq: nfwhalo}
\end{equation}

\begin{equation}
M(r) = 4 \pi \rho_s r_{s}^{3} \ f(x) = M_\mathrm{vir} f(x)/f(c) \ ,
\label{eq: nfwmass}
\end{equation}

\begin{equation}
f(x) \equiv \ln{(1+x)} - \frac{x}{1+x} \ ,
\end{equation}

\begin{equation}
c \equiv r_\mathrm{vir}/r_s \ ,
\label{eq: rvir}
\end{equation}

\begin{equation}
M_\mathrm{vir} = \frac{4 \pi \delta_\mathrm{th}}{3} \Omega_\mathrm{M}
\rho_\mathrm{crit} r_\mathrm{vir}^3,
\end{equation}
where $c$ is the concentration parameter, $M_\mathrm{vir}$ the virial
mass and $r_\mathrm{vir}$ the virial radius\footnote{The virial radius
is defined such that the mean density within $r_\mathrm{vir}$ is
$\delta_\mathrm{th}$ times the mean matter density of the universe
$\bar{\rho} = \Omega_\mathrm{M}
\rho_\mathrm{crit}$. We assume a flat universe with $(\Omega_\mathrm{M},
\Omega_{\Lambda}, h) = (0.3, 0.7, 0.72)$ and $\delta_\mathrm{th} = 337$ (\cite{br+no98}). }.
The model is fully described by two independent parameters, which we
assume to be $c$ and $M_\mathrm{vir}$. A correlation between $c$ and
$M_\mathrm{vir}$ has been found in numerical simulations
(\cite{NFW97,Buletal01,Coletal04}), but we do not use such a relation
since it is affected by a quite large scatter ($\sim 25\%$).

The NFW model is not the only model proposed to fit the results of
numerical simulations. Some authors (\cite{Moore98,Ghigna00}) have
proposed models with a central slope steeper than the NFW one.
However, the difference between these models and the NFW one is very
small for radii larger than 0.5$\%$\,-\,1$\%$ the virial radius and it
is further washed out by the baryonic infall. For these reasons, we
will not consider models different from the NFW one.

\section{Adiabatic compression}
\label{adia}

The present day dark matter halo has a different shape with respect to
the original NFW model since the gravitational collapse of the
baryonic matter, which forms both the bulge and the disc, changes the
overall gravitational potential of the system. The halo structure is
thus modified by the forces of the collapsed baryons depending also on
the angular momentum of the galactic components (baryons and CDM). The
effects of the baryonic infall are treated here following the approach
of the adiabatic compression (\cite{Betal86,FPBF93,MMW98}). If the
disc is assembled slowly, we can assume that the halo responds
adiabatically to the modifications of the gravitational potential and
it remains spherical while contracting. The angular momentum of the
dark matter particles is then conserved and a particle which is
initially at a mean radius $r_i$ ends up at a mean radius $r$ where\,:

\begin{equation}
M_\mathrm{f}(r) \ r = M(r_i) \ r_i \,
\label{eq: acfirst}
\end{equation}
being $M(r_i)$ the initial total mass distribution within $r_i$ and
$M_\mathrm{f}(r)$ the total final mass within $r$. $M_\mathrm{f}(r)$
is the sum of the dark matter inside the initial radius $r_i$ and the
mass contributed by the baryonic components. We thus have\,:

\begin{equation}
M_\mathrm{f}(r) = (1 - m_\mathrm{b}) M(r_i) + M_{bar}(r)
\label{eq: acsecond}
\end{equation}
where $M_{bar}(r)$ is the sum of the contributions from the baryonic
components and $m_\mathrm{b}$ is the fraction of the total mass that
ends up in the visible components. Note that we are implicitly
assuming that the baryons had initially the same mass distribution as
the CDM particles and that those which do not form the luminous units
still remain distributed as the CDM.

For a given rotation curve, $v_\mathrm{c}(R)$, the angular momentum of
the thin disc is\,:

\begin{equation}
J_\mathrm{d} = 2  \pi \int_{0}^{r_\mathrm{vir}}{v_\mathrm{c}(R)
\Sigma_\mathrm{d}^\mathrm{thin}(R) R^2 dR} \
.
\label{eq: jdisc}
\end{equation}
The upper integration limit can be set to infinity since the disc
surface density $\Sigma_\mathrm{d}(R)$ drops exponentially and
$r_\mathrm{vir}$ is much larger than the disc scale\,-\,length. We
assume $J_\mathrm{d}$ to be a fraction $j_\mathrm{d}$ of the initial
angular momentum of the halo, $J$ (i.e. $J_\mathrm{d} = j_\mathrm{d}
J$). $J$ can be expressed in terms of a spin parameter $\lambda$
defined as\,:

\begin{equation}
\lambda \equiv J |E|^{1/2} G^{-1} M_\mathrm{vir}^{-5/2} \ ,
\label{eq: lamdadef}
\end{equation}
with $E$, the total energy of the NFW halo,

\begin{equation}
E = - \frac{G M_\mathrm{vir}^2}{2 r_\mathrm{vir}} f_\mathrm{c} \ .
\label{eq: edef}
\end{equation}
Some simple algebra allows us to rewrite Eq.(\ref{eq: jdisc}) as
(\cite{MMW98})\,:

\begin{equation}
R_\mathrm{d} = \frac{1}{\sqrt{2}} \left (
\frac{j_\mathrm{d}}{m_\mathrm{d}}
\right )
\lambda r_\mathrm{vir} f_\mathrm{c}^{-1/2} f_R(\lambda, c, m_\mathrm{d}, j_\mathrm{d})
\label{eq: rdpar}
\end{equation}
with\,:

\begin{equation}
f_\mathrm{c} = \frac{c}{2} \ \frac{1 - 1/(1+c)^2 - 2 (1+c)^{-1}
\ln{(1+c)}} {[c/(1+c) - \ln{(1+c)}]^2} \ ,
\label{eq: fcdef}
\end{equation}

\begin{equation}
f_R(\lambda, c, m_\mathrm{d}, j_\mathrm{d}) = 2 \left [
\int_{0}^{\infty}{{\rm e}^{-u} u^2
\frac{v_\mathrm{c}(R_\mathrm{d} u)}{v_\mathrm{vir}} du} \right ]^{-1} \ .
\label{eq: frdef}
\end{equation}
In Eqs.(\ref{eq: rdpar}) and (\ref{eq: frdef}), $m_\mathrm{d}$ is the
fraction of the total mass competing to the thin disc, $u \equiv
R/R_\mathrm{d}$ and $v_\mathrm{vir}$ is the total circular velocity at
the virial radius $r_\mathrm{vir}$. It is important to stress that the
rotation curve entering Eq.(\ref{eq: jdisc}) is the total one, i.e. it
is\,:

\begin{equation}
v_\mathrm{c}^{2}(r) = v_\mathrm{c,bar}^{2}(r) + v_\mathrm{c,DM}^{2}(r)
\ ,
\end{equation}
where the first term is simply evaluated given the distribution of the
baryonic components, while the latter is\,:

\begin{equation}
 v_\mathrm{c,DM}^{2}(r) = G [M_\mathrm{f}(r) - M_\mathrm{bar}(r)]/r \ .
\label{eq: vccdm}
\end{equation}

The full set of equations allows one to determine the final
distribution of the DM particles provided that a model for the density
profiles of the baryons has been assigned and the parameters
$(\lambda, m_\mathrm{b}, m_\mathrm{d}, j_\mathrm{d}, c,
M_\mathrm{vir})$ have been fixed. We will see later in Sect. 4 that
these parameters are not all independent since it is possible to find
some physically motivated relations among some of them.

It is worth stressing that there is some debate about the validity of
the adiabatic compression formalism. Jesseit et al. (2002) have found
a substantial agreement between the final dark matter distribution in
numerically simulated haloes and that predicted by the adiabatic
compression approach. On the other hand, this result has been
contradicted on the basis of a set of higher resolution numerical
simulations recently carried out by Gnedin et al. (2004). According to
these authors, the standard adiabatic compression formalism
systematically overpredicts the dark matter density profile in the
inner 5$\%$ of the virial radius. It is worth noting, however, that
only one of the eight simulations considered by Gnedin et al. refers
to a galactic (rather than a cluster) halo. The bottom panel of
Fig.\,4 in their paper shows that the adiabatic compression formalism
overpredict the dark matter density less than $\sim 10\%$ at
$r/r_\mathrm{vir} \sim 0.1$, while the error quickly decreases for
larger values of $r/r_\mathrm{vir}$. This error is much smaller than
the uncertainties we have on the observational quantities so that we
are confident that using the standard adiabatic formalism does not
introduce any bias in the results.

\section{Observational constraints}

We want to select mass models of the Milky Way that are physically
well motivated and are in agreement with observational data. We
discuss in this section the observational constraints used to test
each model.

Only compact baryonic objects can cause microlensing events towards
the Galactic bulge. The number of microlensing events observed towards
the galactic bulge (\cite{Aetal00a,Pop00}) determines the minimum
baryonic mass in the inner Galaxy that can yield the measured value of
the optical depth $\tau$. For an axisymmetric mass density which
decreases moving vertically away from the plane, it is possible to
demonstrate that the minimum baryonic mass is (\cite{BE01})\,:

\begin{equation}
M_{bar}^{min} = \frac{{\rm e} c^2 z_0 \tau}{G}
\label{eq: bebarmass}
\end{equation}
where $z_0 \simeq R_0 \tan{b}$, with $b$ the galactic latitude of the
field target. This is only a lower estimate of the minimum baryonic
mass inside the solar circle and it is quite independent on the
density profile of the baryonic components. Using the latest measured
value of the optical depth towards the Baade window from the analysis
of 52 events in which a clump giant is lensed, $\tau = (2.0 {\pm} 0.4) {\times}
10^{-6}$ (Popowski et al. 2000), we can thus select only those sets of
parameters $(R_0, M_\mathrm{bulge}, \Sigma_{\odot})$ which produce a
baryonic mass within $R_0$ higher than what is predicted by
Eq.(\ref{eq: bebarmass}). Note that we are implicitly assuming that
all the observed microlensing events are due to stellar lenses which
is true only if there is no compact dark matter (such as MACHOs) in
the disc. Should this assumption not to be true, the minimum baryonic
mass should be lowered. However, we stress that this constraint is not
very selective so that changing the value of $M_{bar}^{min}$ does not
affect our results.

Whilst microlensing towards the bulge provides constraints on the
inner Galaxy, satellite dynamics and modeling of the Magellanic Clouds
motion probe the Galaxy mass distribution on a large scale ($\simeq$
50\,-\,100\,kpc). Lin et al. (1995) have used the dynamics of the
Magellanic Clouds to infer the mass of the Milky Way inside 100\,kpc
obtaining $M(r < 100 \ {\rm kpc}) = (5.5 {\pm} 1) {\times} 10^{11} \ M_{\odot}$.
This is in good agreement with the value found by Kochanek (1996)
using escape velocity and motions of the satellite galaxies which give
$(5\,-\,8) {\times} 10^{11} \ M_{\odot}$. Taking into account the different
techniques used and a dependence of the results on the halo modeling,
we follow Dehnen \& Binney (1998) assuming as our constraint\,:

\begin{equation}
M(r < 100 \ {\rm kpc}) = (7.0 {\pm} 2.5) {\times} 10^{11} \ M_{\odot} \ .
\label{eq: testm100}
\end{equation}

A third and quite efficient constraint is given by the Oort's
constants defined as\,:

\begin{equation}
A = \frac{1}{2} \ \left [ \frac{v_\mathrm{c}(R)}{R} -
\frac{dv_\mathrm{c}(R)}{dR} \right ] \ ,
\label{eq: oorta}
\end{equation}

\begin{equation}
B = - \frac{1}{2} \ \left [ \frac{v_\mathrm{c}(R)}{R} +
\frac{dv_\mathrm{c}(R)}{dR}
\right ] \ .
\label{eq: oortb}
\end{equation}
Dehnen \& Binney (1998) have reviewed the estimates of the Oort's
constants present in literature finally proposing the following
values\,:

\begin{equation}
A = (14.5 {\pm} 1.5) \ {\rm km \ s^{-1} kpc^{-1}} \ ,
\label{eq: testoorta}
\end{equation}

\begin{equation}
B = (-12.5 {\pm} 2.0) \ {\rm km \ s^{-1} kpc^{-1}} \ ,
\label{eq: testoortb}
\end{equation}

\begin{equation}
A - B = (27.0 {\pm} 1.5) \ {\rm km \ s^{-1} kpc^{-1}} \ .
\label{eq: testoortab}
\end{equation}
From the definitions of the Oort's constants, it immediately turns out
that $(A - B) R_0 = v_\mathrm{c}(R_0)$. We can thus replace the
constraints on $A - B$ with a constraints on the local circular
velocity\,:

\begin{equation}
v_\mathrm{c}(R_0) = R_0 {\times} (27.0 {\pm} 1.5) \ {\rm km \ s^{-1}} \ .
\label{eq: testvc}
\end{equation}

The vertical force $K_z$ at some height above the plane places a
condition on the local mass distribution. Using K stars as a tracer
population, Kujiken \& Gilmore (1989, 1991) have deduced\,:

\begin{equation}
K_{z, 1.1} \equiv |K_z(R_0, 1.1 \ {\rm kpc}) | = 2 \pi G {\times} (71 {\pm} 6) \
M_{\odot} \ {\rm pc^{-2}} \ .
\label{eq: testgk}
\end{equation}
Formally, this estimate depends on the galactic constants $R_0$ and
$v_\mathrm{c}(R_0)$, but Olling \& Merrifield (2001) have shown that
the result is quite robust against variations in these parameters.
Thus, we can use the previous value to constrain our models.

Finally, another constraint comes from the rotation curve of our
Galaxy. We can reconstruct this quantity from the measurements of the
velocity field. For an axisymmetric galaxy, the radial velocity
relative to the local standard of rest, $v_{r}$, of a circular
orbiting object at galactic coordinates $(l, b)$ and galactocentric
radius $R$ is related to the circular speed by\,:

\begin{equation}
W(R) \equiv \frac{v_{r}}{\sin{l} \cos{b}} =
\frac{R_0}{R} v_\mathrm{c}(R) - v_\mathrm{c}(R_0)
\label{eq: vlsr}
\end{equation}
with the following relation between $R$ and the distance $d$ to the
object\,:

\begin{displaymath}
R^2 = d^2 \cos^2{b} + R_{0}^2 - 2 d R_0 \cos{b} \cos{l} \ .
\end{displaymath}
Several studies are available with measurements of both $d$ and
$v_{r}$ for objects which ought to be on a nearly circular orbits, so
that the Milky Way rotation curve $v_\mathrm{c}(R)$ can be
reconstructed. We use here the data on H\,II regions and molecular
clouds in Brand \& Blitz (1993) and the ones on a sample of classical
Cepheids in the outer disc obtained by Pont et al. (1997). Following
Dehnen \& Binney (1998), we reject objects with either $155^o \le l
\le 205^o$ or $W < 0$ or $d < 1$\,kpc when $v_{r}$ is very likely
dominated by non\,-\,circular motions.

\section{Exploring the parameter space}

To apply the formalism of the adiabatic compression in order to have
the present day halo mass profile, we need the initial halo shape and
the today density profile of the baryonic components. As discussed
above, the baryons have been distributed in the bulge and in the three
sub\,-\,discs according to the observational data.

We have however still an indetermination on the baryonic components
since the Sun distance to the galactic centre $R_0$, the total mass of
the bulge, $M_\mathrm{bulge}$, the local surface density,
$\Sigma_{\odot}$, and the discs scale\,-\,lengths (fixed by the
scaling constant $\kappa$) are known with uncertainties. As a first
step, we consider a grid of models with $R_0$ ranging from 7.0 to
8.5\,kpc in steps of 0.5\,kpc, $M_\mathrm{bulge}$ from 0.80 to 1.80 ${\times}
10^{10} \ M_{\odot}$ with a step of $0.125 {\times} 10^{10} \ M_{\odot}$,
$\Sigma_{\odot}$ from 40 to 56\,$M_{\odot} \ {\rm pc}^{-2}$ in steps
of $1.6 \ M_{\odot} \ {\rm pc}^{-2}$ and $\kappa$ from 0.25 to 0.35 in
step of 0.05. Among these models we select only the ones which pass
the test on the minimum baryonic mass within $R_0$. We stress that
changing the value of $M_{bar}^{min}$ to consider the (quite unlikely)
possibility that some microlensing events are not due to stellar
lenses have only a minor effect on this criterium. The lower is the
value of $M_{bar}^{min}$, the higher is the number of models passing
this preliminary test, but all of the models added by lowering
$M_{bar}^{min}$ will be finally excluded by the selection procedure
described later.

The parameters $(\lambda, m_\mathrm{b}, m_\mathrm{d}, j_\mathrm{d}, c,
M_\mathrm{vir})$ characterize the dark halo and are needed to solve
the adiabatic compression equations. Furthermore, they give scaling
relations between the baryonic components and the total mass
distribution. First, we note that $m_\mathrm{d}$ can be expressed as
function of $m_\mathrm{b}$ as\,:

\begin{equation}
m_\mathrm{d} = \frac{M_\mathrm{thin}}{M_\mathrm{bulge} +
M_\mathrm{thin} + M_\mathrm{thick} + M_\mathrm{ISM}} {\times} m_\mathrm{b}
\equiv f_\mathrm{d} \ m_\mathrm{b}
\ ,
\label{eq: mdmb}
\end{equation}
with $M_\mathrm{thin}$, $M_\mathrm{thick}$ and $M_\mathrm{ISM}$,
respectively, the total mass of the thin, thick and ISM disc. It is
also possible to express the virial mass $M_\mathrm{vir}$ as function
of $M_\mathrm{thin}$ and $m_\mathrm{b}$ simply as (\cite{MMW98})\,:

\begin{equation}
M_\mathrm{vir} = \frac{M_\mathrm{thin}}{m_\mathrm{d}} =
\frac{M_\mathrm{thin}}{f_\mathrm{d} m_\mathrm{b}} \ .
\label{eq: mvirial}
\end{equation}
Equation (\ref{eq: mvirial}) simply states that the final mass of the
system is the same as the initial one. Thus, we are now reduced to
only four parameters, namely $(\lambda, m_\mathrm{b}, j_\mathrm{d},
c)$. We can further reduce the number of parameters observing that
Eqs.(\ref{eq: acfirst}), (\ref{eq: acsecond}) and (\ref{eq: rdpar})
are a system of three independent equations which can be iteratively
solved to determine the initial radius $r_i(r)$, the final mass
$M_\mathrm{f}(r)$ and one of the four parameters, once fixed the
remaining three ones. As we will see later, while there are some hints
about the distribution of the other parameters, little is known about
the value of the concentration $c$. We have thus decided to solve the
set of equations with respect to $(r_i, M_\mathrm{f}, c)$ having fixed
the parameters $(\lambda, m_\mathrm{b}, j_\mathrm{d})$. The equations
are highly non linear and must be solved iteratively\footnote{Mo et
al. (1998) proposed an approximate relation which can be used to
estimate directly $c$ from Eq.(\ref{eq: rdpar}) thus avoiding the
iterative procedure. However, their approximation has been obtained
neglecting the bulge and assuming a single exponential disc instead of
the three sub\,-\,discs we are using. We have checked that their
formula may lead to underestimate strongly the value of $c$.}, so
that, to speed up the calculations, we have imposed a priori that $c$
should be in the range $(5, 25)$ which is a quite conservative
estimate for spiral galaxies similar to the Milky Way (\cite{JVO02}).

To explore in detail the space of parameters, we build, for each model
with given values of $(R_0, M_\mathrm{bulge}, \Sigma_{\odot},
\kappa)$, a set of models individuated by the values of $(\lambda,
m_\mathrm{b}, j_\mathrm{d})$. We briefly explain how we define the
grid. We fix a value for the spin parameter $\lambda$. The
distribution of $\lambda$ for haloes generated in numerical N\,-\,body
simulations is well approximated by a log\,-\,normal distribution with
parameters nearly independent on the cosmological parameters, halo
mass and redshift (\cite{BE87,LK99,Vetal01}). Using the parameters in
Vitvitska et al. (2001), the maximum of the distribution is at
$\lambda = 0.035$ while there is a 90$\%$ probability that $\lambda$
is in the range $(0.02, 0.10)$. We thus let $\lambda$ change in this
range in steps of 0.01. Next, we have to fix a value for
$m_\mathrm{b}$. This parameter is poorly constrained since we only
know that it cannot be larger than the universal baryon fraction
$\Omega_\mathrm{b}/\Omega_\mathrm{M}$. This latter has been inferred
by observations involving completely different physical processes
(\cite{turn01}). The power spectrum of matter inhomogeneities from
observations of large-scale structure is sensitive to
$\Omega_\mathrm{b}/\Omega_\mathrm{M}$; the Two Degree Field Galaxy
Redshift Survey has reported a value of $0.15 {\pm} 0.07$
(\cite{per+al01}). Measurements of the angular power spectrum of the
CMBR also provide a very significant estimate. The combined analysis
in Jaffe et al. (2001) of several data set gives
$\Omega_\mathrm{b}/\Omega_\mathrm{M} = 0.186^{+0.010}_{-0.008}$. We
thus let $m_\mathrm{b}$ change from 0.01 to 0.20 in steps of 0.01.
Finally, we have to fix $j_\mathrm{d}$. This parameter also is not
constrained either theoretically or by numerical simulations. On one
hand, it is reasonable to assume $j_\mathrm{d}
= m_\mathrm{d}$ and indeed this seems to be necessary to fit spiral galaxies
rotation curves (\cite{MMW98}). On the other hand, numerical
simulations have found $j_\mathrm{d}/m_\mathrm{d}$ significantly less
than unity. We have varied this parameter from 0.005 to $m_\mathrm{d}$
in steps of 0.01.

The grid we build in this way is quite detailed. For each given $(R_0,
M_\mathrm{bulge}, \Sigma_{\odot}, \kappa)$ the total number of sets
$(\lambda, m_\mathrm{b}, j_\mathrm{d})$ is $\sim 800 - 1200$
(depending on the value of $m_\mathrm{b}$), so that the parameter
space is indeed checked intensively.

To select among this large number of models we use, for each model
parameterization, a multi\,-\,step procedure. First, we solve
iteratively the set of Eqs.(\ref{eq: acfirst}), (\ref{eq: acsecond})
and (\ref{eq: rdpar}) so that we have the full halo mass distribution.
We are thus able to estimate $M_\mathrm{f}(r = 100 \ {\rm kpc})$, the
total mass inside $r = 100$\,kpc. A model passes to the next step only
if this value is consistent with the estimate in Eq.(\ref{eq:
testm100}). In the second step, we compare the local circular velocity
$v_\mathrm{c}(R_0)$ to the constraint in Eq.(\ref{eq: testvc})
rejecting the model if there is no agreement. Next, the Oort's
constant are evaluated and the model is retained if the values of $A$
and $B$ are in agreement with the values given by Eqs.(\ref{eq:
testoorta}) and (\ref{eq: testoortb}). Then, we compute $K_{z, 1.1}$
and accept the model if the resulting value is in agreement with the
estimate reported in Eq.(\ref{eq: testgk}). For the surviving models,
we estimate the $\chi^2$ defined as\,:

\begin{equation}
\chi^2 = \frac{1}{N} \sum \frac{v_{c,data}^{2}(R_i) - v_{c,model}^{2}(R_i)}{\sigma_{i}^{2}}
\label{eq: defchi}
\end{equation}
where $\sigma_i$ is the error on the $i$\,-\,th measurements and the
sum runs over the $N$ data points. Note that we compute the $\chi^2$
as the last step of our selection procedure when all the model
parameters are fixed so that we do not change their values in order to
minimize the $\chi^2$. We deem a model as acceptable if $\chi^2 <
1.33$. For our data set of $N = 115$ entries, this corresponds to a
99.7$\%$ confidence level.

Actually, as stressed by Olling \& Merrifield (2001), selecting among
different models on the basis of the $\chi^2$ value is not a
statistically correct procedure because the errors on both
$v_\mathrm{c}(R)$ and the Galactic constants, which enter the estimate
of $v_\mathrm{c}(R)$ through Eq.(\ref{eq: vlsr}), are not normal. This
means that a high value of $\chi^2$ for a given model might be a
consequence of an intrinsically wrong model or of the not normal
origin of the errors. Furthermore, the data of Pont et al. (1997)  on
the radial velocities of the outer disc classical Cepheids are given
without any uncertainties, so that, for these data points, $\sigma_i$
is only determined by the propagation of the errors on the galactic
constants and it is thus underestimated. So we have decided to still
retain the $\chi^2$\,-\,test a selection criterium, but we also
consider the median statistics. As shown in Gott et al.\,(2001; see
also Avelino et al. 2002, Chen \& Ratra 2003, Sereno 2003), median
statistics provide a powerful alternative to $\chi^2$ likelihood
methods. Fewer assumptions about the data are needed. A proper median
statistics assume that {\it (i)} experimental results are
statistically independent; {\it (ii)} there are no systematic effects.
Statistical errors are not required to be either known or gaussianly
distributed. Since our analysis is based on not normal errors,
performing a test without using the errors themselves turns out to be
a very conservative approach. Furthermore, median statistics is also
less vulnerable to the presence of bad data and outliers.

To compute the likelihood of a particular set of parameters, we count
how many data points are above or below each model prediction and
compute the binomial likelihoods. Given a binomial distribution, if we
perform $N$ measurements, the probability of obtaining $k$ of them
above the median is given by
\begin{equation}
\label{bino}
P(k)=\frac{2^{-N}N!}{k!(N-k)!} \ .
\end{equation}
We count how many of the 115 experimental points are above the
expected velocity rotation curve, and retain a model if the number of
overestimates is between 43 and 72. Given the distribution in
Eq.~(\ref{bino}), the probability that the median of 115 sorted
entries falls in this range is 99.73\%.

It is worth to note that our final results do not depend on the
sequence of the tests.

\section{Analysis of the selection criteria}

Before presenting the results of our analysis, it is interesting to
investigate at what stage in our selection procedure certain types of
models are excluded, i.e. we want to study the impact of each
criterium on the parameter space. To this aim, for a given set of
galactic parameters $(R_0, M_\mathrm{bulge}, \Sigma_{\odot}, \kappa)$,
we first select all the models with $c$ in the range $(5, 25)$ and
then apply to this set of models the selection criteria introduced in
the previous section separately. The main results of this analysis are
presented in Figs.\,1\,--\,5 and discussed below as regard the two
parameters entering the adiabatic compression equations, i.e. $c$ and
$m_\mathrm{b}$.

\begin{figure}
\resizebox{8.5cm}{!}{\includegraphics{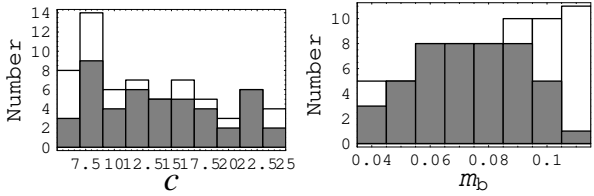}}
\caption{Histogram of the number distribution of $c$ and of $m_\mathrm{b}$ the for the models
with $(R_0, M_\mathrm{bulge}, \Sigma_{\odot}, \kappa) = (8.5, 1.0, 54,
0.3)$ and $c$ between 5 and 25. $R_0$ is in kpc, $M_\mathrm{bulge}$ in
$10^{10} \ M_{\odot}$, $\Sigma_{\odot}$ in $M_{\odot}/pc^2$. The
shaded histogram refers to the models passing the test on $M(r < 100 \
{\rm kpc})$.}
\end{figure}

\begin{figure}
\resizebox{8.5cm}{!}{\includegraphics{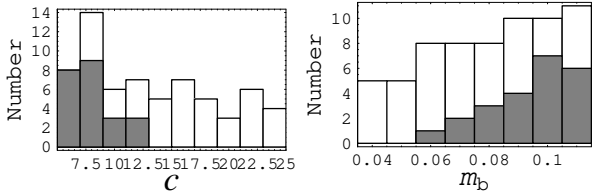}}
\caption{Same as Fig.\,1 for the test on $v_\mathrm{c}(R_0)$.}
\end{figure}

\begin{itemize}

\item{{\it Constraint on $M(r < 100 \ {\rm kpc})$.} As Fig.\,1 shows, the application
of this criterium tends to flatten the histogram of the $c$ values,
being however slightly more effective in cutting out models with
values of $c$ in the tails of the distribution. On the contrary, this
cut tends to suppress the high end of the $m_\mathrm{b}$ histogram,
only retaining those models with smaller $m_\mathrm{b}$. This latter
result may be qualitatively explained considering Eq.(\ref{eq:
mvirial}) which shows that the higher is $m_\mathrm{b}$, the lower is
$M_\mathrm{vir}$ and thus $M(r < 100 \ {\rm kpc})$.}

\item{{\it Constraint on $v_\mathrm{c}(R_0)$.} Fig.\,2 shows the impact of the test on the
local circular velocity. High values of $c$ are excluded by this
constraint. As regard the $m_\mathrm{b}$ histogram, this selection
criterium turns out to be almost orthogonal to the previous one since
now low values of $m_\mathrm{b}$ are clearly disfavoured.}

\item{{\it Constraint on the Oort constants.} This test turns out to be a
sort of compromise between the two previous ones. On one hand, it is
quite effective in excluding models with very low values of $c$ (look
at the lowest bin), as it is (with much less efficiency) for the
constraint on $M(r < 100 \ {\rm kpc})$. On the other hand, it cuts
away the low end of the $m_\mathrm{b}$ histogram in the same way as
the selection criterium based on the value of $v_\mathrm{c}(R_0$) do.
This is a quite important result since it shows that the eventual
exclusion of this constraint does not alter the final results of the
multi-step selection procedure. The presence of local fluctuations in
the ISM disc density strongly affects the derivatives of the
gravitational potential thus leading to possible errors in the
evaluation of the Oort constants for a given set of galactic
parameters. One could thus argue that it should be better to not use
the Oort constants as a selection criterium, but only their difference
and hence the local circular velocity. However, the use of a third
order polynomial interpolation of the measured ISM disc density
alleviates this problem so that we are confident that the estimated
values of $A, B$ are not corrupted. Furthermore, as we have observed
comparing Fig.\,3 with Figs.\,1\,--\,2, applying only the constraint
on $A, B$ gives results that are consistent with those obtained using
the two constraints on $M(r < 100 \ {\rm kpc})$ and
$v_\mathrm{c}(R_0)$ so that any systematic error in the estimate of
the Oort constants is washed out in our multi-step procedure.}

\begin{figure}
\resizebox{8.5cm}{!}{\includegraphics{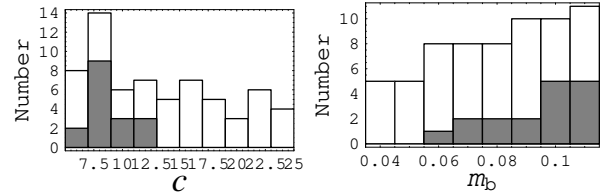}}
\caption{Same as Fig.\,1 for the test on the Oort constants.}
\end{figure}

\begin{figure}
\resizebox{8.5cm}{!}{\includegraphics{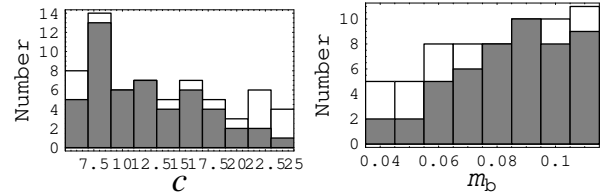}}
\caption{Same as Fig.\,1 for the test on $K_{z,1.1}$.}
\end{figure}

\item{{\it Constraint on $K_{z,1.1}$.} This test works excluding models with values of
$c$ and $m_\mathrm{b}$ in the tails of the distribution. This
constraint is very effective when selecting among different galactic
parameters leading to reject models with $\Sigma_{\odot} < 49 \
M_{\odot}/pc^2$. Actually, the lower is $\Sigma_{\odot}$, the higher
is the percentage of models excluded by the test on $M(r < 100 \ {\rm
kpc})$ or by that on the Oort constants. This is a clear evidence
against models with low values of the local surface density in
agreement with what the results of the application of the constraint
on $K_{z,1.1}$ claim.}

\item{{\it Constraint on $\chi^2$.} The application of this constraint allows
to flatten the histogram of the $c$ values lowering the peak in
Fig.\,5 for low values. However, the high end of the histogram is
erased thus suggesting that models with very high values of $c$ are
not able to fit the Milky Way rotation curve. This is not an
unexpected result since both N\,-\,body simulations and fitting of the
adiabatically compressed NFW model to external galaxies show that
small values of $c$ are best suited to describe galactic dark haloes
(\cite{JVO02}). As concern $m_\mathrm{b}$, the constraint on the
$\chi^2$ works as that on the local circular velocity selecting models
with high values of this parameter. This is a reasonable result since
both constraints are related to the same physical quantity, i.e. the
rotation curve. The median statistics works the same way as the
$\chi^2$\,-\,test, but it is more stringent.}

\end{itemize}

\begin{figure}
\resizebox{8.5cm}{!}{\includegraphics{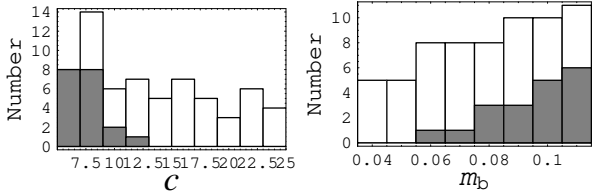}}
\caption{Same as Fig.\,1 for the test on $\chi^2$ value.}
\end{figure}

As a final remark, we want to stress that the observational data we
have reviewed in Sect.\,3 are ``consensus" values with ``consensus"
errors so that these latter can not be treated as ``statistical"
uncertainties. This is the reason why we have decided to adopt a
multi-step selection procedure instead of the usual $\chi^2$
minimization technique based on the definition of a $\chi^2$ entering
all the constraints at the same time. Our filtering approach and the
discussion presented in this section allow one to avoid all the
problems connected with the statistics of not normal errors and makes
it possible to understand how the results could vary changing one of
the constraints. Actually, the procedure we have implemented is quite
robust since the different constraints select different regions of the
parameter space thus allowing to narrow the ranges for both the
adiabatic compression parameters and the galactic constants.

\section{Results}

The selection procedure we have employed is quite efficient allowing
us to reject the most of the models. Since we have used two
alternative test as last constraint (the $\chi^2$ value or the median
statistics), we define two samples. Sample A contains those models
passing all the selection criteria and having $\chi^2 < 1.33$, while
Sample B is made out by the models passing the test on the median
statistics. The final number of models is 116 in the Sample A and 34
in the Sample B from an initial set of $\sim 10^6$. Actually, it turns
out that the Sample B is a subset of Sample A, i.e. all the models in
Sample B belong to Sample A too. This is expected since the median
statistics is a more restrictive test than the $\chi^2$ analysis.

The main results are shown in Figs.(6-9) and summarized in Tables 1
and 2. Note that the results obtained from the two samples are in
perfect agreement so that we will not discuss them separately.  We
stress that the final samples of models may not be treated using the
usual statistical methods since our selection procedure is based on
constraints on different observable that are ``consensus" values with
``consensus" errors. That is why we do not report as best estimate of
the parameters their mean values, but the medians which is a more
conservative approach. Because of this, the quoted 68$\%$ (95$\%$)
range must not be considered as the $1 -
\sigma$ ($2 - \sigma$) confidence limit, but it is simply the range which the 68$\%$
(95$\%$) of the values are within. With this caveat in mind, we
discuss below the distribution of the various model parameters.

\begin{table}
\caption{Median values, $68\%$ and $95\%$ regions for the model parameters
of sample A. The total halo mass $M_\mathrm{vir}$ is expressed in
units
 of $10^{11} \ M_{\odot}$.}
\begin{center}
\begin{tabular}{|c|c|c|c|}
\hline
Parameter & Median & $68\% $ range & $95\%$ range \\
\hline
$c$ & 6.89 & 5.49 -- 9.45 & 5.04 -- 11.88 \\ $M_\mathrm{vir}$ & 8.68 &
7.53
-- 10.20 & 6.87 -- 16.60 \\ $m_\mathrm{b}$ & 0.08 & 0.07 -- 0.09 & 0.04 -- 0.10
\\ $\lambda$ & 0.06 & 0.04 -- 0.09 & 0.03 -- 0.10 \\ $\lambda'$ &
0.029 &  0.026 -- 0.035 & 0.018 -- 0.042 \\
\hline
\end{tabular}
\end{center}
\end{table}

\begin{table}
\caption{Median values, $68\%$ and $95\%$ regions for the model parameters
of sample B. The total halo mass $M_\mathrm{vir}$ is expressed in
units
 of $10^{11} \ M_{\odot}$}
\begin{center}
\begin{tabular}{|c|c|c|c|}
\hline
Parameter & Median & $68\% $ range & $95\%$ range \\
\hline
$c$ & 6.48 & 5.49 -- 7.26 & 5.08 -- 9.75 \\ $M_\mathrm{vir}$ & 8.44 &
7.58 -- 9.39 & 7.03 -- 11.16 \\ $m_\mathrm{b}$ & 0.08 & 0.07 -- 0.09 &
0.06 -- 0.09 \\ $\lambda$ & 0.06 & 0.04 -- 0.09 & 0.03 -- 0.10 \\
$\lambda'$ & 0.029 & 0.027 -- 0.032 & 0.023 -- 0.035 \\
\hline
\end{tabular}
\end{center}
\end{table}

\begin{enumerate}

\item{Figure 6 shows the distribution of the values of the concentration
parameter $c$. Small values are clearly favoured so that the median
value for Sample A (Sample B) comes out to be $c = 6.89$ $(6.44)$. It
is worth stressing that the values of $c$ we get are lower than
expected, but are not unrealistic. Jimenez et al. (2003) have fitted
the rotation curves of 400 spiral galaxies by modelling them with an
exponential disc and a dark halo obtained by adiabatically compressing
the NFW profile. Their Fig. 1 shows the distribution of the
concentration $c$ vs the total mass $M_\mathrm{vir}$. There are indeed
a lot of galaxies with values of $c$ in the same range as the one
found here.}

\begin{figure}
\resizebox{8.5cm}{!}{\includegraphics{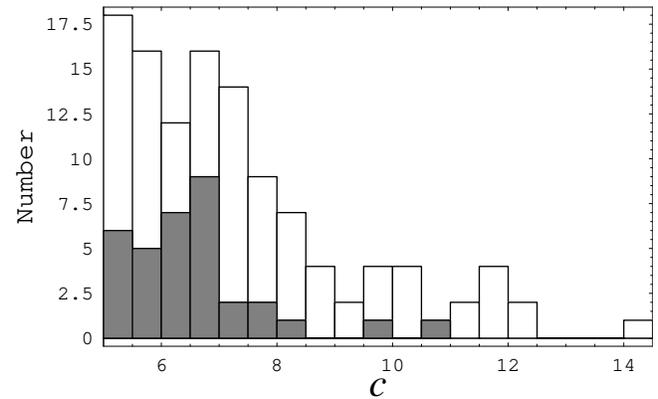}}
\caption{Histogram of the number distribution of the concentration
parameter $c$ for the models in Sample A and Sample B (shaded
histogram).}
\end{figure}

\begin{figure}
\resizebox{8.5cm}{!}{\includegraphics{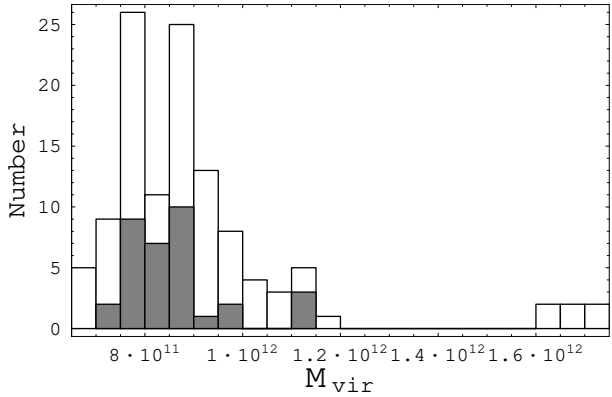}}
\caption{Same as Fig. 6 for the total halo mass $M_\mathrm{vir}$.}
\end{figure}

\item{In Fig. 7, we plot the distribution of the values of the total
mass $M_\mathrm{vir}$ for the final set of models. The median value
for Sample A (Sample B) is $8.68 \ (8.44) {\times} 10^{11} \ M_{\odot}$, and
the 95$\%$ confidence interval ranges from 5.04 (5.08)  to $6.74 \
(9.75) {\times} 10^{11} \ M_{\odot}$. These values are still in agreement
with the most recent measurement of the Milky Way total mass estimated
to be $1.9_{-1.7}^{+3.6} {\times} 10^{12} \ M_{\odot}$ (\cite{WE99}). It is
noteworthy, however, that the value given by Evans \& Wilkinson is
higher than usual being the most of the previous estimates lower than
$10^{12} \ M_{\odot}$ (see, e.g., Fig.\,3 in Zaritsky 1999). Sample A
also contains few models with very high values of $M_\mathrm{vir}$.
The value of $M_\mathrm{vir}$ we have obtained may thus be considered
quite reasonable. It is also possible to see that a clear correlation
exists between $c$ and $M_\mathrm{vir}$, with the lower values of the
concentration parameter $c$ giving rise to the highest values of the
total mass. This correlation may reflect the existence of some
covariance among the model parameters in the analysis procedure.}

\item{The distribution of $m_\mathrm{b}$ is shown in Fig. 8. The median value
is $0.08$ for both Sample A and B, while the $95\%$ range turns out to
be 0.07\,-\,0.09. According to some authors
(\cite{MMW98,JVO02,KZS01}), $m_\mathrm{b}$ should be written as
$\varepsilon {\times}
\Omega_\mathrm{b}/\Omega_\mathrm{M}$ with $\varepsilon$ indicating the efficiency of the
transfer of baryons from the initial halo to the bulge and disc. If
this were correct, our results should mean $\varepsilon \simeq 37 - 48
\%$ for $\Omega_\mathrm{b}/\Omega_\mathrm{M} = 0.186$. The efficiency
is not one and this rises the problem of understanding where the
missing baryons are.}

\item{The median value of the spin parameter $\lambda$ is $0.06$ for
both Sample A and B and the $95\%$ range is $0.04 - 0.09$. Models with
low spin are erased by the selection procedure. We do not report any
result for $j_\mathrm{d}$ since this parameter is found to be
degenerate with $\lambda$ and $m_\mathrm{d}$, i.e., for fixed values
of $(m_\mathrm{b}, R_0, M_\mathrm{bulge},
\Sigma_{\odot}, \kappa)$, models having the same value of $\lambda' =
\lambda j_\mathrm{d}/ m_\mathrm{d}$ give rise to the same present day halo mass profile.
This result is not unexpected since in the only equation containing
$\lambda$ and $j_\mathrm{d}$, Eq.(\ref{eq: rdpar}), these two
parameters appear only in $\lambda'$. The median value for this latter
is $\lambda' = 0.029$ with a $95\%$ range going from 0.018 to 0.042
for Sample A and from 0.023 to 0.035 for Sample B.}

\begin{figure}
\resizebox{8.5cm}{!}{\includegraphics{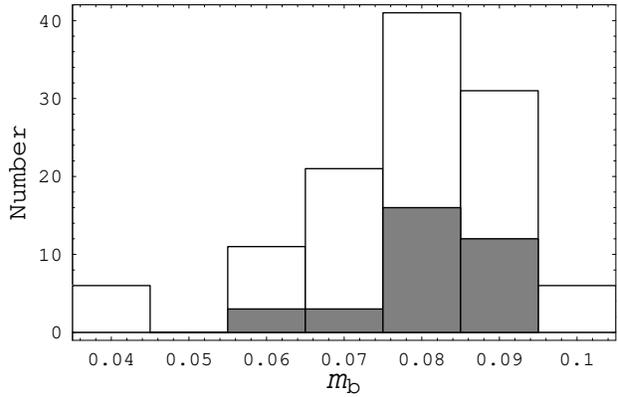}}
\caption{Same as Fig. 6 for $m_\mathrm{b}$.}
\end{figure}

\begin{figure}
\resizebox{8.5cm}{!}{\includegraphics{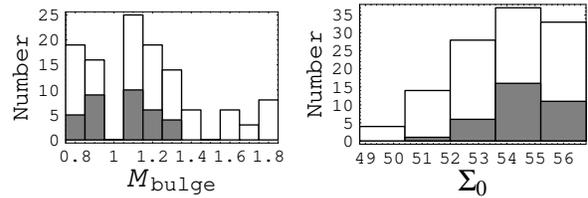}}
\caption{Histograms of the number distribution of models according to the values of
$M_\mathrm{bulge}$ ($10^{10} \ M_{\odot}$, left panel) and
$\Sigma_{\odot}$ ($M_{\odot} \ {\rm pc^{-2}}$, right panel). Shaded
regions refer to Sample B.}
\end{figure}

\item{Figure 9 shows the histogram of the number of models according to the values
of $M_\mathrm{bulge}$ and $\Sigma_{\odot}$. We can draw some
interesting limits. First, we observe that we are not able to
significantly constraint the bulge total mass. Indeed, the median
value of the distribution turns out to be $M_\mathrm{bulge} = 1.0 {\times}
10^{10} \ M_{\odot}$, but the 95$\%$ range for Sample A does not
reject any model. It is worth noting, however, that values of
$M_\mathrm{bulge}$ in the range $0.8 - 1.4 {\times} 10^{10} \ M_{\odot}$
seems to be favoured. This is confirmed by analysing Sample B that
excludes models with $M_\mathrm{bulge} > 1.4 {\times} 10^{10} \ M_{\odot}$. A
similar analysis leads to the following constraints on the disc local
surface density\,:

\begin{displaymath}
49 \ M_{\odot} \ {\rm pc}^{-2} \le \Sigma_{\odot} \le 56 \ M_{\odot} \
{\rm pc}^{-2} \
\end{displaymath}
with $54 \ M_{\odot} \ {\rm pc^{-2}}$ as median value. If we subtract
the contribution of the ISM disc surface density, we get for the
stellar disc $\Sigma_{\star} \simeq 40  \ M_{\odot} \ {\rm pc^{-2}}$
in agreement with the consensus value ($\Sigma_{\star} = 35 {\pm} 10 \
M_{\odot} \ {\rm pc^{-2}}$) proposed by Olling \& Merrifield (2001).
Only 6 out of 116 models in Sample A have $R_0 = 8.0$\,kpc, while in
all the other cases it is $R_0 = 8.5$\,kpc so that we may safely
consider $R_0 = 8.5$\,kpc as our final estimate of this galactic
constant. This conclusion is strengthened by noting that all the
models in Sample B have $R_0 = 8.5$\,kpc. This value is somewhat
higher than expected since most recent estimates predict values in the
range 7.0 - 8.0 kpc (\cite{R93,OM00}). However, the estimate of $R_0$
is somewhat model dependent since its value is linked to the halo
flattening. For instance, Olling \& Merrifield (2001) have found that
it is possible to build galaxy models with $R_0 \stackrel{>}{\sim}
7.0$\,kpc if the halo is close to spherical which indeed is our case.
Finally, there are no models with $\kappa \ne 0.30$ so that we
conclude that the discs scale\,-\,length is $R_\mathrm{d} = 0.30 {\times} R_0
= 2.55$\,kpc. This is lower than the fiducial value (3.5 kpc)
suggested by Binney \& Tremaine (1987) and often adopted in many Milky
Way disc modelling. However, we note that a value of $R_\mathrm{d}
\simeq 2.5$ is favoured by both star count models
(\cite{Robin92,OL93}) and integrated NIR luminosity profile
(\cite{Freund,BGS97}).}

\end{enumerate}

\begin{figure}
\resizebox{8.5cm}{!}{\includegraphics{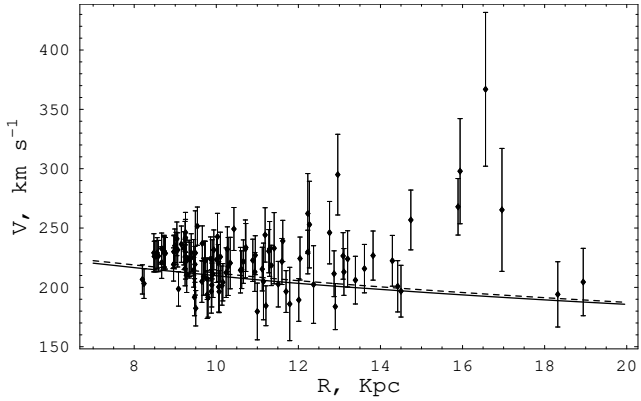}}
\caption{Rotation velocity curve in the Galactic plane for the models
$\left\{ \Sigma_{\odot}, M_\mathrm{bulge}, c, m_\mathrm{b},
j_\mathrm{d},
\lambda
\right\}=
\left\{ 54.4 M_{\odot}{\rm pc}^{-2}, 1.05 {\times} 10^{10}M_{\odot} , 5.55, 0.08, 0.035,
0.04 \right\}$ (full line) and $\left\{\Sigma_{\odot},
M_\mathrm{bulge}, c, m_\mathrm{b}, j_\mathrm{d},\lambda \right\}=
\left\{ 56.0 M_{\odot}{\rm pc}^{-2}, 9.25 {\times} 10^{9}M_{\odot} , 6.56, 0.09, 0.045,
0.04 \right\}$ (dashed line).}
\label{FigVelRot}
\end{figure}

Finally, we wish to comment again on the use of the $\chi^2$ as a
selection criterium. Actually, it is difficult to say whether a
certain value of $\chi^2$ for a given model is due to some systematic
error (as having neglected the triaxial structure of the bulge and the
flattening of the halo) or to a problem with the parameter values or
to the not normal nature of the estimated errors on the rotation
curve. However, the agreement among the results obtained using both
the $\chi^2$ analysis and the median statistics (that turns out to be
a more selective criterium) is a convincing evidence that the excluded
models have been rejected as a consequence of physical problems,
either due to the need for a more careful description of the inner
Galaxy or to an intrinsically wrong combinations of the parameters
$(\lambda, m_\mathrm{b}, j_\mathrm{d}, c)$. This makes us confident in
the results. As an example, in Fig.~\ref{FigVelRot} we plot the
rotation curve in the Galactic plane for two models, the one with the
lower value of the $\chi^2$ ($\left\{ \Sigma_{\odot},
M_\mathrm{bulge}, c, m_\mathrm{b}, j_\mathrm{d},\lambda
\right\}=
\left\{ 54.4 M_{\odot}{\rm pc}^{-2}, 1.05 {\times} 10^{10}M_{\odot} , 5.55, 0.08, 0.035,
0.04 \right\}$) and the one with median test value equal to 58
($\left\{ \Sigma_{\odot}, M_\mathrm{bulge}, c, m_\mathrm{b},
j_\mathrm{d},\lambda
\right\}=
\left\{ 56.0 M_{\odot}{\rm pc}^{-2}, 9.25 {\times} 10^{9}M_{\odot} , 6.56, 0.09, 0.045,
0.04 \right\}$). As we can see, the difference in the rotation curve
between the two models is really negligible. We remind that each set
of model with passes the selection procedure has the same statistical
weight.

\section{Conclusions}

Modelling Milky Way is one of the classical tasks of astronomy. In
this paper, we have applied the adiabatic compression method in the
framework of the CDM structure formation to build models of the Galaxy
in agreement with the observational data and, at the same time, well
motivated by numerical simulations of galaxy formation. Exploring in
very detail the parameter space, we have finally selected a set of
models which has allowed us to draw some interesting constraints on
$(i)$ the galactic parameters, namely the Sun distance to the Galactic
centre $R_0$, the total mass of the bulge $M_\mathrm{bulge}$, the disc
local surface density $\Sigma_{\odot}$ and the disc scale\,-\,length
$R_\mathrm{d}$, $(ii)$ the parameters entering the adiabatic
compression, i.e. the spin parameter $\lambda$, the fraction of the
mass in baryons $m_\mathrm{b}$ and the thin disc contribution
$j_\mathrm{d}$ to the total angular momentum and $(iii)$ the
parameters describing the initial halo profile (i.e. $c$ and
$M_\mathrm{vir}$).

It should be interesting to compare our results with previous ones in
literature, but a straight comparison is not possible because of the
different approaches followed in the analyses. Klypin et al. (2002)
have used the adiabatic compression theory to build Milky Way models
which are in agreement with observational data starting from a NFW
halo. Even if the final aim is the same as our one, the approach used
is radically different since they give a value for the concentration
of the initial NFW halo and then find by trial and errors the values
of the disc and bulge parameters so that the model is in agreement
with the data. In our approach, the visible components are fixed from
the beginning and the halo parameters are determined later. Klypin et
al. (2002) finally examine only four models. Their favoured one is
radically different from our ones since it has a longer disc
scale\,-\,length (3.5 vs 2.55 kpc) and a higher concentration (12 vs.
6.48). However, it is noteworthy  that their Model A2 seems to be more
in agreement with our results, having $c = 5$ and $M_\mathrm{vir} =
7.1 {\times} 10^{11} \ M_{\odot}$, and the authors find that also this model
is able to fit the data. We stress, however, that most of the
disagreements between our results and those of Klypin et al. (2002)
may derive from a different way of selecting the models compatible
with the observational constraints.

A somewhat surprising result of our analysis is the quite small values
of the concentration parameter $c$ that turns out to be smaller than
11.88 (9.75) at 95$\%$ for Sample A (Sample B). Using the relation
between $M_\mathrm{vir}$ and $c$ in Bullock et al. (2001), for values
of $M_\mathrm{vir}$ in the 95$\%$ range reported in Table 1, one
should expect $c$ in the range 14.78 -- 16.04. Considering a $\sim
25\%$ scatter (\cite{Coletal04}), our median $c$ is more than
2$\sigma$ smaller than the above lower limit. Moreover, Fig.\,6 shows
that none of the models have such high values of $c$ so that we may
conclude that the quoted relation $c - M_\mathrm{vir}$ found in
numerical simulations is not verified by our observationally selected
Milky Way mass models. We have verified that this result is
independent on our choice of cosmological parameters by repeating
analysis for an Einstein-de Sitter model of universe
($\Omega_\mathrm{M}=1,\Omega_\Lambda=0$).

Although surprising, this result is not fully unexpected. Fitting
adiabatically compressed NFW models to the rotation curve of a sample
of 400 spiral galaxies, Jimenez et al. (2003) have determined the
values of $c$ and $M_\mathrm{vir}$ for these haloes. Looking at their
Fig.\,1, one sees that there are a lot of galaxies with values of
$M_\mathrm{vir}$ in the range determined by us for the Milky Way. For
these galaxies, the concentration parameter turns out to be of the
same order as those obtained here so that the relation $c -
M_\mathrm{vir}$ is not satisfied for these galaxies too. We are thus
confident that the disagreement we have found for the selected Milky
Way models is not a result of the our procedure, but a possible
shortcoming either of the numerical simulations or of the $\Lambda$CDM
paradigm itself.

The procedure presented may be extended to other spiral galaxies. A
possible target is the Andromeda galaxy whose visible components can
be modelled in detail thanks to the available photometric and
cinematic data. It is also interesting to apply our procedure to low
surface brightness (LSB) galaxies. Modelling in detail the visible
components in these galaxies is much difficult because of the paucity
of the data, but these systems are probably dark matter dominated so
that uncertainties in the baryonic components should not affect
systematically the main results. Moreover, combining the constraints
from many galaxies will allow reducing the systematics connected with
the disc modelling. To this aim, a sample of LSB with high resolution
rotation curve should be ideal since it allows to decouple the disc
contribution from the halo one (\cite{dBB02}).

The analysis carried out may be further refined in order to
investigate whether the constraints on the parameters are affected by
some simplifying hypotheses. We have implicitly assumed that there is
no exchange of angular momentum between the dark matter particles and
the baryons infalling in the disc and bulge. An approximate analytical
approach to this problem has been developed (\cite{KZS01}) and it
should be thus interesting to repeat our analysis by including this
effect. However, the details of how the baryons are transferred from
the halo to the disc depend on the physics of star formation and
supernovae explosion and some kind of energy feedback is needed to
eliminate the so\,-\,called cooling catastrophe (\cite{Baletal01}). In
order to investigate these effects, numerical simulations are needed,
but the physics of the process may be also described with
semi\,-\,analytical modelling (\cite{Cole,vdB01}). These methods are
best suited to be included in our procedure, thus making it possible
to put stronger constraints both on the adiabatic compression
parameters $(\lambda, m_\mathrm{b}, j_\mathrm{d})$ and on the initial
NFW halo ones $(c, M_\mathrm{vir})$.

We would like to conclude with a general consideration. The aim of
this paper has been to build a Milky Way mass model in agreement with
the observational data and motivated by some physical background. The
simple adiabatic compression formalism has allowed us to take into
account analytically the effect of baryonic infall in a way which is
consistent with more complex numerical simulations. We have finally
found that such a model indeed exists although there is a disagreement
with the predicted concentration parameter. In our opinion, this could
be considered as an indirect evidence suggesting that the hierarchical
CDM scenario of galaxy formation is essentially correct since its
predictions are in agreement with the data on the best studied galaxy,
the Milky Way. We are thus confident that solutions to the problems of
this cosmological model do not require modifications of dark matter
properties, but have to be searched in astrophysical phenomena.

\begin{acknowledgements}
We acknowlege an anonymous referee for having pointed out a major
error in our procedure and for his comments that have helped us to
significantly improve the paper. We warmly thank Walter Dehnen for
having furnished us his code for computing the gravitational potential
of the thin and thick discs. It is a pleasure to thank also Shude Mao
and Yogesh Joshi for some useful comments.
\end{acknowledgements}

\end{document}